\documentclass[runningheads]{llncs}

\usepackage{cite}
\usepackage[colorlinks=true,
    linkcolor=blue,
    urlcolor=blue,citecolor=blue]{hyperref}
\usepackage{amsmath}
\usepackage{multirow}
\usepackage{array}
\usepackage{amssymb}
\usepackage{booktabs}
\usepackage{stfloats}
\usepackage{url}
\usepackage{color}
\usepackage{xspace}
\usepackage{bm}

\usepackage{algorithm}
\usepackage{algpseudocode}

\usepackage[misc]{ifsym}
\usepackage[T1]{fontenc}

\usepackage{graphicx}

\begin{document}

\title{ASCON: Anatomy-aware Supervised Contrastive Learning Framework for Low-dose CT Denoising}

\titlerunning{Anatomy-aware Low-dose CT Denoising}

\author{Zhihao Chen\inst{1} \and
Qi Gao\inst{1} \and Yi Zhang\inst{2} \and
Hongming Shan\inst{1}\textsuperscript{(\Letter)}}

\authorrunning{Z. Chen et al.}

\institute{Institute of Science and Technology for Brain-Inspired Intelligence, Fudan University, Shanghai, China \\
\email{hmshan@fudan.edu.cn}
\and
School of Cyber Science and Engineering, Sichuan University, Chengdu, China
}

\maketitle              

\newcommand{\etal}{\textit{et al}.\xspace}
\newcommand{\ie}{\textit{i}.\textit{e}.\xspace}
\newcommand{\eg}{\textit{e}.\textit{g}.\xspace}
\newcommand{\tabincell}[1]{\begin{tabular}[l]{@{}l@{}} #1\end{tabular}}

\newcommand{\mat}[1]{\boldsymbol{#1}}
\newcommand{\vct}[1]{\boldsymbol{#1}}

\newcommand{\std}[1]{\tiny #1}

\newcommand{\modelname}{ASCON\xspace}
\newcommand{\contrablock}{MAC-Net\xspace}
\newcommand{\generator}{ESAU-Net\xspace}
\newcommand{\globalblock}{GPN\xspace}
\newcommand{\localblock}{LPC\xspace}
\newcommand{\contra}{GLC\xspace}

\algnewcommand{\CommentS}[1]{\State \textbf{Comment:} \emph{#1}}

\begin{abstract}
While various deep learning methods have been proposed for low-dose computed tomography (CT) denoising, most of them leverage the normal-dose CT images as the ground-truth to supervise the denoising process. 
These methods typically ignore the inherent correlation within a single CT image, especially the anatomical semantics of human tissues, and lack the interpretability on the denoising process.  
In this paper, we propose a novel \textbf{A}natomy-aware \textbf{S}upervised \textbf{CON}trastive learning framework, termed  \modelname, which can explore the anatomical semantics for low-dose CT denoising while providing anatomical interpretability.
The proposed \modelname consists of two novel designs: an efficient self-attention-based U-Net~(\generator) and a multi-scale anatomical contrastive network~(\contrablock). 
First, to better capture global-local interactions and adapt to the high-resolution input, an efficient \generator is introduced by using a channel-wise self-attention mechanism.
Second, \contrablock incorporates a patch-wise non-contrastive module to capture inherent anatomical information and a pixel-wise contrastive module to maintain intrinsic anatomical consistency.
Extensive experimental results on two public low-dose CT denoising datasets demonstrate superior performance of \modelname over state-of-the-art models. Remarkably, our \modelname provides anatomical interpretability for low-dose CT denoising for the first time. Source code is available at \url{https://github.com/hao1635/ASCON}.
\end{abstract}

\section{Introdcution}
With the success of deep learning in the field of computer vision and image processing, many deep learning-based methods have been proposed and achieved promising results in low-dose CT (LDCT) denoising~\cite{redcnn,liang2020edcnn,shan2019competitive,shan20183,wgan-vgg,huang2021gan,geng2021content,cocodiff,corediff}. 
Typically, they employ a supervised learning setting, which involves a set of image pairs, LDCT images and their normal-dose CT (NDCT) counterparts. These methods typically use a pixel-level loss (\eg mean squared error or MSE), which can cause over-smoothing problems. 

To address this issue, a few studies~\cite{shan2019competitive,wgan-vgg} used a structural similarity (SSIM) loss or a perceptual loss~\cite{johnson2016perceptual}. 
However, they all perform in a sample-to-sample manner and ignore the inherent anatomical semantics, which could blur details in areas with low noise levels. 
Previous studies have shown that the level of noise in CT images varies depending on the type of tissues~\cite{mussmann2021organ}; see an example in Fig.~\ref{mean_std} in Supplementary Materials. 
Therefore, it is crucial to characterize the anatomical semantics for effectively denoising diverse tissues. 

In this paper, we focus on taking advantage of the inherent anatomical semantics in LDCT denoising from a contrastive learning perspective~\cite{grill2020bootstrap,yun2022patch,yan2022sam}.
To this end, we propose a novel \textbf{A}natomy-aware \textbf{S}upervised \textbf{CON}trastive learning framework (\modelname), which consists of two novel designs: an efficient self-attention-based U-Net~(\generator) and a multi-scale anatomical contrastive network~(\contrablock). 
First, to ensure that \contrablock can effectively extract anatomical information, diverse global-local contexts and a larger input size are necessary. However, operations on full-size CT images with self-attention are computationally unachievable due to potential GPU memory limitations~\cite{petit2021u}. To address this limitation, we propose an \generator that utilizes a channel-wise self-attention mechanism~\cite{ronneberger2015u,zamir2022restormer,chen2023lit} 
which can efficiently capture both local and global contexts by computing cross-covariance across feature channels.

Second, to exploit inherent anatomical semantics, we present the \contrablock that employs a disentangled U-shaped architecture~\cite{yan2022sam} to produce global and local representations. Globally, a patch-wise non-contrastive module is designed to select neighboring patches with similar semantic context as positive samples and align the same patches selected in denoised CT and NDCT which share the same anatomical information, using an optimization method similar to the BYOL method~\cite{grill2020bootstrap}. This is motivated by the prior knowledge that adjacent patches often share common semantic contexts~\cite{yun2022patch}.
Locally, to further improve the anatomical consistency between denoised CT and NDCT, we introduce a pixel-wise contrastive module with a hard negative sampling strategy~\cite{robinson2020contrastive}, which randomly selects negative samples from the pixels with high similarity around the positive sample within a certain distance. Then we use a local InfoNCE loss~\cite{oord2018representation} to pull the positive pairs and push the negative pairs.

Our contributions are summarized as follows.
1) We propose a novel \modelname framework to explore inherent anatomical information in LDCT denoising, which is important  to provide interpretability for LDCT denoising.
2) To better explore anatomical semantics in \contrablock, we design an \generator, which utilizes a channel-wise self-attention mechanism to capture both local and global contexts.
3) We propose a \contrablock that employs a disentangled U-shaped architecture and incorporates both global non-contrastive and local contrastive modules. This enables the exploitation of inherent anatomical semantics at the patch level, as well as improving anatomical consistency at the pixel level.
4) Extensive experimental results demonstrate that our \modelname  outperforms other state-of-the-art methods, and provides anatomical interpretability for LDCT denoising.

\begin{figure}[t]
\centering
\includegraphics[width=0.9\linewidth]{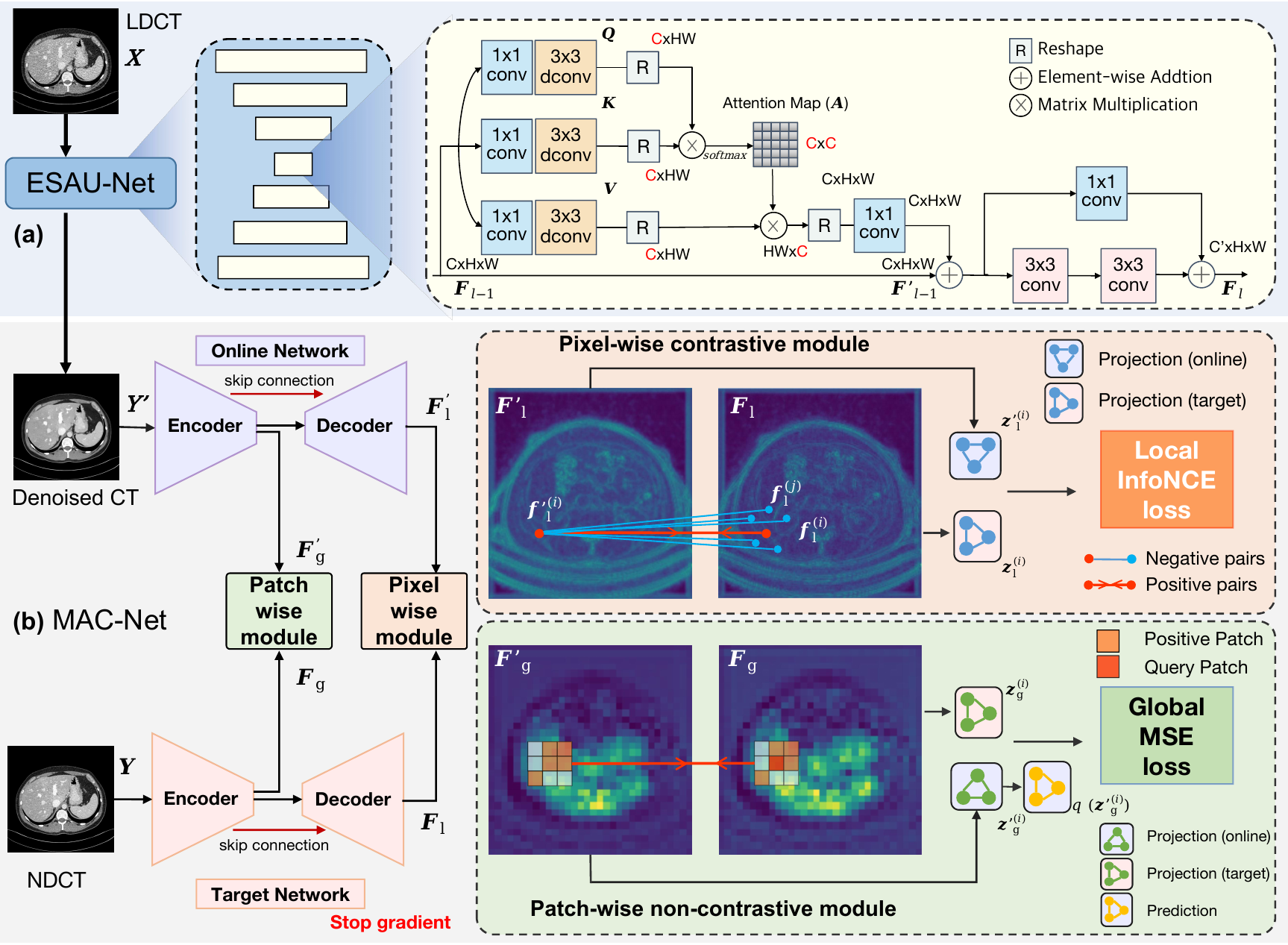}
\caption{Overview of our proposed \modelname. (a) Efficient self-attention-based U-Net~(\generator); and (b) multi-scale anatomical contrastive network~(\contrablock).}
\label{network:ascon}
\end{figure}

\section{Methodology}
\subsection{Overview of the proposed \modelname} 
Fig.~\ref{network:ascon} presents the overview of the proposed \modelname, which consists of two novel components: \generator and \contrablock. First, given an LDCT image, $\mat{X} \in \mathbb{R}^{1  \times H \times W}$, where $H\times W$ denotes the image size. $\mat{X}$ is passed through the \generator to capture both global and local contexts using a channel-wise self-attention mechanism and obtain a denoised CT image $\mat{Y}^{\prime} \in \mathbb{R}^{1 \times H \times W}$.

Then, to explore inherent anatomical semantics and remain inherent anatomical consistency, the denoised CT $\mat{Y}^{\prime}$ and NDCT $\mat{Y}$ are passed to the \contrablock to compute a global MSE loss $\mathcal{L}_\mathrm{{global}}$ in a patch-wise non-contrastive module and a local infoNCE loss $\mathcal{L}_\mathrm{{local}}$ in a pixel-wise contrastive module. During training, we use an alternate learning strategy to optimize \generator and \contrablock separately, which is similar to GAN-based methods~\cite{isola2017image}. Please refer to Alg.~\ref{alternate_learning} in Supplementary Materials for a detailed optimization.

\subsection{Efficient Self-attention-based U-Net}
\label{generator}
To better leverage anatomical semantic information in \contrablock and adapt to the high-resolution input, we design the \generator that can capture both local and global contexts during denoising. 
Different from previous works that only use self-attention in the coarsest level~\cite{petit2021u}, we incorporate a channel-wise self-attention mechanism~\cite{chen2023lit,zamir2022restormer} at each up-sampling and down-sampling level in the U-Net~\cite{ronneberger2015u} and add an identity mapping in each level, as shown in Fig.~\ref{network:ascon}(a). 

Specifically, in each level, given the feature map $\mat{F}_{l-1}$ as the input, we first apply a $1\times1$ convolution and a $3\times3$ depth-wise convolution to aggregate channel-wise contents and generate query ($\mat{Q}$), key ($\mat{K}$), and value ($\mat{V}$) followed by a reshape operation, where $\mat{Q} \in \mathbb{R}^{C\times HW}$, $\mat{K} \in \mathbb{R}^{C\times HW }$, and $\mat{V} \in \mathbb{R}^{C\times HW }$ (see Fig.~\ref{network:ascon}(a)).
Then, a channel-wise attention map $\mat{A}\in \mathbb{R}^{C\times C}$ is generated through a dot-product operation by the reshaped query and key, which is more efficient than the regular attention map of size ${HW\times HW}$~\cite{dosovitskiy2020image}, especially for high-resolution input. Overall, the process is deﬁned as
\begin{align}
\tiny
\operatorname{Attention}(\mat{F})&=w(\mat{V}^{T}\mat{A}) 
=w(\mat{V}^{T} \cdot \operatorname{Softmax}({\mat{K}\mat{Q}^{T}}/{\alpha})),
\end{align}
where $w(\cdot)$ first reshapes the matrix back to the original size $C\times H \times W$ and then performs  $1\times1$ convolution; $\alpha$ is a learnable parameter to scale the magnitude of the dot product of $\mat{K}$ and $\mat{Q}$. 
We use multi-head attention similar to the standard multi-head self-attention mechanism~\cite{dosovitskiy2020image}. The output of the channel-wise self-attention is represented as: ${\mat{F}^{\prime}_{l-1}}=\operatorname{Attention}(\mat{F}_{l-1})+\mat{F}_{l-1}$. Finally, the output $\mat{F}_{l}$ of each level is defined as:
$\mat{F}_{l}=\operatorname{Conv}(\mat{F}^{\prime}_{l-1})+\operatorname{Iden}(\mat{F}_{l-1})$,
where $\operatorname{Conv}(\cdot)$ is a two-layer convolution and $\operatorname{Iden}(\cdot)$
 is an identity mapping using a $1\times1$ convolution; refer to Fig.~\ref{network_esau_dunet}(a) for the details of \generator.

\subsection{Multi-scale Anatomical Contrastive Network}
\label{MSM}
\noindent\textbf{Overview of \contrablock.}\quad
The goal of our \contrablock is to exploit anatomical semantics and maintain anatomical embedding consistency, 
First, a disentangled U-shaped architecture~\cite{ronneberger2015u} is utilized to learn global representation $\mat{F}_\mathrm{g}$ $\in\mathbb{R}^{512 \times \frac{H}{16} \times \frac{W}{16}}$ after four down-sampling layers, and learn local representation $\mat{F}_\mathrm{l}$$\in\mathbb{R}^{64 \times H \times W}$ by removing the last output layer. And we cut the connection between the coarsest feature and its upper level to make $\mat{F}_\mathrm{g}$ and $\mat{F}_\mathrm{l}$ more independent~\cite{yan2022sam} (see Fig. \ref{network_esau_dunet}(b)). 
The online network and the target network, both using the same architecture above, handle denoised CT $\mat{Y}^{\prime}$ and NDCT $\mat{Y}$, respectively, with $\mat{F}^{\prime}_\mathrm{g}$ and $\mat{F}^{\prime}_\mathrm{l}$ generated by the online network, and $\mat{F}_\mathrm{g}$ and $\mat{F}_\mathrm{l}$ generated by the target network (see Fig.~\ref{network:ascon}(b)).
The parameters of the target network are an exponential moving average of the parameters in the online network, following the previous works~\cite{he2020momentum,grill2020bootstrap}. Next, a patch-wise non-contrastive module uses $\mat{F}_\mathrm{g}$ and $\mat{F}^{\prime}_\mathrm{g}$ to compute a global MSE loss $\mathcal{L}_\mathrm{global}$, while a pixel-wise contrastive module uses $\mat{F}_\mathrm{l}$ and $\mat{F}^{\prime}_\mathrm{l}$ to compute a local infoNCE loss $\mathcal{L}_\mathrm{local}$. Let us describe these two loss functions specifically.

\noindent\textbf{Patch-wise non-contrastive module.}\quad
To better learn anatomical representations, we introduce a patch-wise non-contrastive module, also shown in Fig.~\ref{network:ascon}(b). Specifically, for each pixel ${{\vct{f}}_\mathrm{g}^{(i)}} \in \mathbb{R}^{512}$ in the ${\mat{F}}_\mathrm{g}$ where $i\in\{1,2, \ldots, \frac{HW}{256}\}$
is the index of the pixel location, it can be considered as a patch due to the expanded receptive field achieved through a sequence of convolutions and down-sampling operations~\cite{park2020contrastive}. To identify positive patch indices, we adopt a neighboring positive matching strategy~\cite{yun2022patch}, 
assuming that a semantically similar patch ${{\vct{f}}_\mathrm{g}^{(j)}}$ exists in the vicinity of the query patch ${{\vct{f}}_\mathrm{g}^{(i)}}$, as neighboring patches often share a semantic context with the query.  
We empirically consider a set of 8 neighboring patches.
To sample patches with similar semantics around the query patch ${{\vct{f}}_\mathrm{g}^{(i)}}$, we measure the semantic closeness between the query patch ${{\vct{f}}_\mathrm{g}^{(i)}}$ and its neighboring patches ${{\vct{f}}_\mathrm{g}^{(j)}}$ using the cosine similarity, which is formulated as 
\begin{equation}
\small
s(i, j)=({{\vct{f}}_\mathrm{g}^{(i)}})^{\top} ({{\vct{f}}_\mathrm{g}^{(j)}})/\|{{\vct{f}}_\mathrm{g}^{(i)}}\|_{2}\|{{\vct{f}}_\mathrm{g}^{(j)}}\|_{2}.
\label{eq:similarity}
\end{equation}

We then select the top-4 positive patches $\{{{\vct{f}}_\mathrm{g}^{(j)}}\}_{j \in \mathcal{P}^{(i)}}$ based on $s(i, j)$,  where $\mathcal{P}^{(i)}$ is a set of selected patches (\ie, $|\mathcal{P}^{(i)}|=4$). To obtain patch-level features $\mat{{g}}^{(i)} \in \mathbb{R}^{512}$ for each patch ${{\vct{f}}_\mathrm{g}^{(i)}}$ and its positive neighbors, we aggregate their features using global average pooling (GAP) in the patch dimension. For the local representation of ${{\vct{f}^{\prime}}_\mathrm{g}^{(i)}}$, we select positive patches as same as $\mathcal{P}^{(i)}$, \ie,~$\{{{\vct{f}^{\prime}}_\mathrm{g}^{(j)}}\}_{j \in \mathcal{P}^{(i)}}$. Formally, 
\begin{equation}
\small
\vct{g}^{(i)}:=\operatorname{GAP}({{\vct{f}}_\mathrm{g}^{(i)}},\{{{{\vct{f}}_\mathrm{g}^{(j)}}\}_{j \in \mathcal{P}^{(i)}}}),\quad 
{{\mat{{g}}^{\prime}}^{(i)}}:=\operatorname{GAP}({{\vct{f}^{\prime}}_\mathrm{g}^{(i)}},\{{{{\vct{f}^{\prime}}_\mathrm{g}^{(j)}}\}_{j \in \mathcal{P}^{(i)}}}).
\end{equation}

From the patch-level features, the online network outputs a projection ${{\vct{z}_\mathrm{g}^{\prime}}^{(i)}}={{p^{\prime}_\mathrm{g}}({{\vct{g}^{\prime}}^{(i)})}}$ and a prediction ${q^{\prime}}{({\vct{z}_\mathrm{g}^{\prime}}^{(i)})}$ while target network outputs the target projection ${\vct{z}_\mathrm{g}}^{(i)}={{p_\mathrm{g}}({{\vct{g}}^{(i)})}}$. The projection and prediction are both multi-layer perceptron (MLP). Finally, we compute the global MSE loss between the normalized prediction and target projection~\cite{grill2020bootstrap},
\begin{equation}
    \mathcal{L}_\mathrm{global} =\sum\nolimits_{i\in{\mathcal{N}_{\mathrm{pos}}^\mathrm{g}}}\big\|{q^{\prime}}({\vct{z}_\mathrm{g}^{\prime}}^{(i)})-{\vct{z}_\mathrm{g}}^{(i)}\big\|_{2}^{2}=\sum\nolimits_{i\in{\mathcal{N}_{\mathrm{pos}}^\mathrm{g}}}2-2 \cdot \tfrac{\langle {q^{\prime}}({\vct{z}_\mathrm{g}^{\prime}}^{(i)}), {\vct{z}_\mathrm{g}}^{(i)}\rangle}{\|{q^{\prime}}({\vct{z}_\mathrm{g}^{\prime}}^{(i)})\|_{2} \cdot\|{\vct{z}_\mathrm{g}}^{(i)}\|_{2}},
\end{equation}
where ${\mathcal{N}_{\mathrm{pos}}^\mathrm{g}}$ is the indices set of positive samples in the patch-level embedding.

\noindent\textbf{Pixel-wise contrastive module.}\quad
In this module, we aim to improve anatomical consistency between the denoised CT and NDCT using a local InfoNCE loss~\cite{oord2018representation} (see Fig.~\ref{network:ascon}(b)). First, for a query ${{\vct{f}^{\prime}_\mathrm{l}}^{(i)}} \in \mathbb{R}^{64}$ in the $\mat{F}^{\prime}_\mathrm{l}$ and its positive sample ${{\vct{f}}_\mathrm{l}^{(i)}} \in \mathbb{R}^{64}$ in the ${\mat{F}}_\mathrm{l}$ ($i\in\{1,2, \ldots, HW\}$ is the location index), we use a hard negative sampling strategy~\cite{robinson2020contrastive} to select ``diffcult'' negative samples with high probability, which enforces the model to learn more from the fine-grained details. Specifically, candidate negative samples are randomly sampled from ${\mat{F}}_\mathrm{l}$ as long as their distance from ${{\vct{f}}_\mathrm{l}^{(i)}}$ is less than $m$ pixels ($m$=7). We also use cosine similarity in Eq.~\eqref{eq:similarity} to select a set of semantically closest pixels, \ie~$\{{{\vct{f}}_\mathrm{l}^{(j)}}\}_{j \in \mathcal{N}_\mathrm{neg}^{(i)}}$. Then we concatenate ${{\vct{f}^{\prime}}_\mathrm{l}^{(i)}}, {{\vct{f}}_\mathrm{l}^{(i)}},$ and $\{{{{\vct{f}}_\mathrm{l}^{(j)}}\}_{j \in \mathcal{N}_\mathrm{neg}^{(i)}}}$ and map them to a $K$-dimensional vector ($K$=256) through a two-layer MLP, obtaining 
The local InfoNCE loss in the pixel level is deﬁned as
\begin{align}
\tiny
\mathcal{L}_\mathrm{local}=-\sum\nolimits_{i\in{\mathcal{N}^\mathrm{l}_{\mathrm{pos}}}} \log \tfrac{\exp \left({{\vct{v}^{\prime}}_\mathrm{l}^{(i)}} \cdot {{\vct{v}}_\mathrm{l}^{(i)}} / \tau\right)}{\exp \left({{\vct{v}^{\prime}}_\mathrm{l}^{(i)}} \cdot {{\vct{v}}_\mathrm{l}^{(i)}} / \tau\right)+\sum_{j=1}^{|{\mathcal{N}_{\mathrm{neg}}^{(i)}}|} \exp \left({{\vct{v}^{\prime}}_\mathrm{l}^{(i)}} \cdot {{\vct{v}}_\mathrm{l}^{(j)}} / \tau\right)},
\end{align}
where ${\mathcal{N}_{\mathrm{pos}}^\mathrm{l}}$ is the indices set of positive samples in the pixel level. ${{\vct{v}^{\prime}}_\mathrm{l}^{(i)}}$, ${{\vct{v}}_\mathrm{l}}^{(i)}$, and ${{\vct{v}}_\mathrm{l}^{(j)}}\in\mathbb{R}^{256}$ are the query, positive, and negative sample in ${{\vct{z}}_\mathrm{l}^{(i)}}$, respectively. 
\subsection{Total Loss Function}
The final loss is defined as 
$\mathcal{L}=\mathcal{L}_\mathrm{global}+\mathcal{L}_\mathrm{local}+\lambda\mathcal{L}_\mathrm{pixel},$
where $\mathcal{L}_\mathrm{pixel}$ consists of two common supervised losses: MSE and SSIM, defined as $\mathcal{L}_\mathrm{pixel}=\mathcal{L}_\mathrm{MSE}+\mathcal{L}_\mathrm{SSIM}$. $\lambda$ is empirically set to $10$.

\section{Experiments}
\subsection{Dataset and Implementation Details}
 We use two publicly available low-dose CT datasets released by the NIH AAPM-Mayo Clinic Low-Dose CT Grand Challenge in 2016~\cite{mccollough2017low} and lately released in 2020~\cite{moen2021low}, denoted as Mayo-2016 and Mayo-2020, respectively. There is no overlap between the two datasets. Mayo-2016 includes normal-dose abdominal CT images of 10 anonymous patients and corresponding simulated quarter-dose CT images. Mayo-2020 provides the abdominal CT image data of 100 patients with 25\% of the normal dose, and we randomly choose 20 patients for our experiments.
 
For the Mayo-2016, we choose 4800 pairs of $512\times512$ images from 8 patients for the training and 1136 pairs from the rest 2 patients as the test set. For the Mayo-2020, we employ 9600 image pairs with a size of $256\times256$ from randomly selected 16 patients for training and 580 pairs of $512\times512$ images from rest 4 patients for testing. The use of large-size training is to adapt our \contrablock to exploit inherent semantic information. The default sampling hyper-parameters for Mayo-2016 are ${|\mathcal{N}_{\mathrm{pos}}^\mathrm{l}|=32}$, ${|\mathcal{N}_{\mathrm{pos}}^\mathrm{g}|=512}$, ${|\mathcal{N}_{\mathrm{neg}}^{(i)}}|=24$, while ${|\mathcal{N}_{\mathrm{pos}}^\mathrm{l}|=16}$, ${\mathcal{|N}_{\mathrm{pos}}^\mathrm{g}|=256}$, ${|\mathcal{N}_{\mathrm{neg}}^{(i)}}|=24$ for Mayo-2020. We use a binary function to filter the background while selecting queries in \contrablock. For the training strategy, we employ a window of [$-$1000, 2000] HU. We train our network for 100 epochs on 2 NVIDIA GeForce RTX 3090, and use the AdamW optimizer~\cite{loshchilov2017decoupled} with the momentum parameters $\beta_{1}=0.9$, $\beta_{2}=0.99$ and the weight decay of $1.0\times10^{-9}$. We initialize the learning rate as $1.0\times 10^{-4}$, gradually reduced to $1.0\times 10^{-6}$ with the cosine annealing~\cite{loshchilov2016sgdr}. Since \contrablock is only implemented during training, the testing time of \modelname is close to most of the compared methods.

\begin{table}[t]
\caption{Performance comparison on the Mayo-2016  and Mayo-2020 datasets in terms of PSNR [dB], RMSE [$\times 10^{-2}$], and SSIM [\%]}
\label{Results_comparison}
\centering
\resizebox*{0.97\linewidth}{.32\columnwidth}{
\begin{tabular*}{1\linewidth}{@{\extracolsep{\fill}}rcrrrcrrr}
\toprule[1.5pt]
  \multirow{2}{*}{\textbf{Methods}} && \multicolumn{3}{c}{\textbf{Mayo-2016}}         && \multicolumn{3}{c}{\textbf{Mayo-2020}} \\
\cline{3-5}  \cline{7-9}
&& PSNR$\uparrow$ &  RMSE$\downarrow$      & SSIM$\uparrow$ && PSNR$\uparrow$   &   RMSE$\downarrow$   & SSIM$\uparrow$ \\
\midrule
{\scriptsize{U-Net~\cite{ronneberger2015u}}}   &&  44.13\std{$\pm${1.19}} & 0.64\std{$\pm${0.12}} & 97.38\std{$\pm${1.09}}    &&     47.67\std{$\pm${1.64}}  & 0.43\std{$\pm${0.09}} & 99.19\std{$\pm${0.23}}  \\

{\scriptsize{RED-CNN~\cite{redcnn}}}   && 44.23\std{$\pm${1.26}} & 0.62\std{$\pm${0.09}} & 97.34\std{$\pm${0.86}}   && 48.05\std{$\pm${2.14}}     & 0.41\std{$\pm${0.11}} & 99.28\std{$\pm${0.18}}    \\

{\scriptsize{WGAN-VGG}~\cite{wgan-vgg}}  && 42.49\std{$\pm${1.28}}  & 0.76\std{$\pm${0.12}}  & 96.16\std{$\pm${1.30}} &&  46.88\std{$\pm${1.81}}     & 0.46\std{$\pm${0.10}}   & 98.15\std{$\pm${0.20}}    \\

{\scriptsize{EDCNN~\cite{liang2020edcnn}}}  &&  43.14\std{$\pm${1.27}}  & 0.70\std{$\pm${0.11}}  &  96.45\std{$\pm${1.36}}   &&  47.90\std{$\pm${1.27}}    &  0.41\std{$\pm${0.08}}   & 99.14\std{$\pm${0.17}}    \\

{\scriptsize{DU-GAN~\cite{huang2021gan}}}  && 43.06\std{$\pm${1.22}}  & 0.71\std{$\pm${0.10}}  &  96.34\std{$\pm${1.12}}   && 47.21\std{$\pm${1.52}}   & 0.44\std{$\pm${0.10}}  & 99.00\std{$\pm${0.21}}    \\

{\scriptsize{CNCL}~\cite{geng2021content}}      && 43.06\std{$\pm${1.07}}  & 0.71\std{$\pm${0.10}}  & 96.68\std{$\pm${1.11}}    && 45.63\std{$\pm${1.34}}   & 0.53\std{$\pm${0.11}}  &  98.92\std{$\pm${0.59}}     \\

\hline
{\scriptsize{\generator~(ours)}} &&   \underline{44.38}\std{$\pm${1.26}} & \underline{0.61}\std{$\pm${0.09}} &   \underline{97.47}\std{$\pm${0.87}}        && \underline{48.31}\std{$\pm${1.87}}   & \underline{0.40}\std{$\pm${0.12}}  & \underline{99.30}\std{$\pm${0.18}}       \\

{\scriptsize{\textbf{\modelname~(ours)}}} && {\textbf{44.48}}\std{$\pm${1.32}} & {\textbf{0.60}}\std{$\pm${0.10}} & {\textbf{97.49}}\std{$\pm${0.86}}   && \textbf{48.84}\std{$\pm${1.68}} & \textbf{0.37}\std{$\pm${0.11}} &   \textbf{99.32}\std{$\pm${0.18}}      \\ 
\bottomrule[1.5pt]
\end{tabular*}
}
\end{table}

\subsection{Performance Comparisons}
\noindent\textbf{Quantitative evaluations.}\quad
We use three widely-used metrics such as peak signal-to-noise ratio (PSNR), root-mean-square error (RMSE), and SSIM.  Table~\ref{Results_comparison} presents the testing results on Mayo-2016 and Mayo-2020 datasets. We compare our methods with 5 state-of-the-art methods, including RED-CNN~\cite{redcnn}, WGAN-VGG~\cite{wgan-vgg}, EDCNN~\cite{liang2020edcnn}, DU-GAN~\cite{huang2021gan}, and CNCL~\cite{geng2021content}. Table~\ref{Results_comparison} shows that our \generator with \contrablock achieves the best performance on both the Mayo-2016 and the Mayo-2020 datasets. Compared to the \generator, \modelname further improves the PSNR by up to 0.54 dB on Mayo-2020, which demonstrates the effectiveness of the proposed \contrablock and the importance of the inherent anatomical semantics during CT denoising. We also compute the contrast-to-noise ratio (CNR) to assess the detectability of a selected area of low-contrast lesion and our \modelname achieves the best CNR in Fig.~\ref{CNR}.
\begin{figure}[t]
\centering
\includegraphics[width=0.92\linewidth]{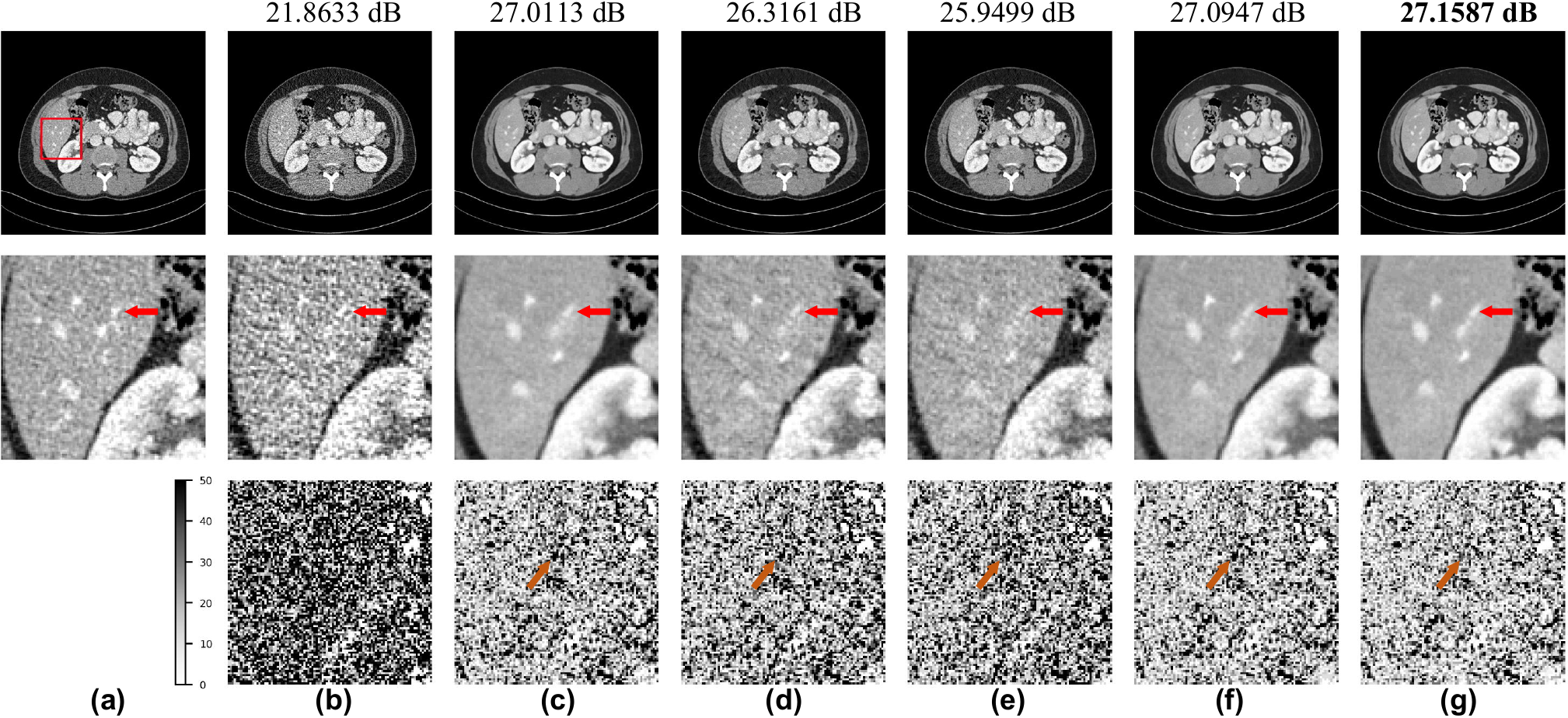}
\caption{Transverse CT images and corresponding difference images from the Mayo-2016 dataset: (a) NDCT; (b) LDCT; (c) RED-CNN~\cite{redcnn}; (d) EDCNN~\cite{liang2020edcnn}; (e) DU-GAN~\cite{huang2021gan}; (f) \generator~(ours); and (g) \modelname~(ours). The display window is [-160, 240] HU.
}
\label{test_results_spatial}
\end{figure}

\noindent\textbf{Qualitative evaluations.}\quad
Fig.~\ref{test_results_spatial} presents qualitative results of three representative methods and our \generator with \contrablock on Mayo-2016. 
Although \modelname and RED-CNN produce visually similar results in low-contrast areas after denoising. However, RED-CNN  results in blurred edges between different tissues, such as the liver and blood vessels, while \modelname smoothed the noise and maintained the sharp edges. They are marked by arrows in the regions-of-interest images. We further visualize the corresponding difference images between NDCT and the generated images by our method as well as other methods as shown in the third row of Fig.~\ref{test_results_spatial}. Note that our \modelname removes more noise components than other methods; refer to Fig.~\ref{spatial_2020} for extra qualitative results on Mayo-2020.

\begin{figure}[t]
\centering
\includegraphics[width=0.89\linewidth]{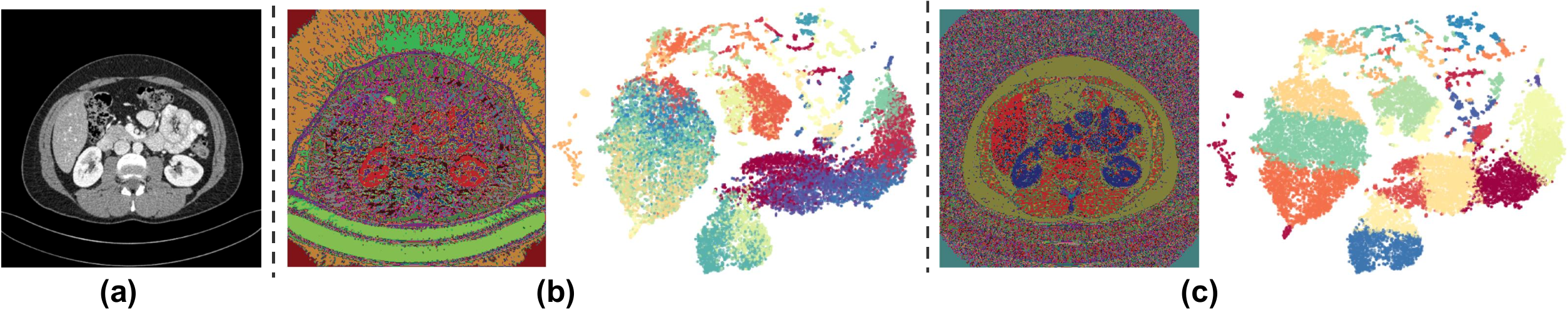}
\caption{Visualization of inherent semantics; (a) NDCT; (b) clustering and t-SNE results of \modelname w/o \contrablock; and (c) clustering and t-SNE results of \modelname. }
\label{interpretability}
\end{figure}

\noindent\textbf{Visualization of inherent semantics.}\quad
To demonstrate that our \contrablock can exploit inherent anatomical semantics of CT images during denoising, we select the features before the last layer in \modelname without \contrablock and \modelname from Mayo-2016. Then we cluster these two feature maps respectively using a K-means algorithm and visualize them in the original dimension, and finally visualize the clustering representations using t-SNE, as shown in Fig.~\ref{interpretability}. Note that \modelname produces a result similar to organ semantic segmentation after clustering and the intra-class distribution is more compact, as well as the inter-class separation is more obvious. To the best of our knowledge, this is the first time that anatomical semantic information has been demonstrated in a CT denoising task, providing interpretability to the field of medical image reconstruction. 

\begin{table}[t]
\centering
\renewcommand\arraystretch{0.8}
\caption{Ablation results of Mayo-2020 on the different types of loss functions.}

\label{loss_functions}
\begin{tabular*}{0.9\linewidth}{@{\extracolsep{\fill}}lrrr}
\toprule[1.5pt]
\scriptsize{Loss}     & \scriptsize{PSNR}$\uparrow$ & \scriptsize{RMSE}$\downarrow$ & \scriptsize{SSIM}$\uparrow$\\
 \midrule
$\mathcal{L}_\mathrm{MSE}$     &  48.34\std{$\pm${2.22}}  & 0.40\std{$\pm${0.11}}  & 99.27\std{$\pm${0.18}}  \\

$\mathcal{L}_\mathrm{MSE}+ \mathcal{L}_\mathrm{Perceptual}$ &   47.83\std{$\pm${1.99}}    &   0.42\std{$\pm${0.10}}   &  99.13\std{$\pm${0.19}}    \\

$\mathcal{L}_\mathrm{MSE}+ \mathcal{L}_\mathrm{SSIM}$  &  48.31\std{$\pm${1.87}}   &  0.40\std{$\pm${0.12}}   & 99.30\std{$\pm${0.18}}  \\

$\mathcal{L}_\mathrm{MSE}+ \mathcal{L}_\mathrm{SSIM}+\mathcal{L}_\mathrm{global}$  &  48.58\std{$\pm${2.12}}    &  0.39\std{$\pm${0.10}}  &  99.31\std{$\pm${0.17}}  \\ 
$\mathcal{L}_\mathrm{MSE}+\mathcal{L}_\mathrm{SSIM}+\mathcal{L}_\mathrm{local}$   & 48.48\std{$\pm${2.37}}   &  0.38\std{$\pm${0.11}}  & 99.31\std{$\pm${0.18}} \\
$\mathcal{L}_\mathrm{MSE}+ \mathcal{L}_\mathrm{SSIM}+$ $\mathcal{L}_\mathrm{local} +\mathcal{L}_\mathrm{global}$   &  \textbf{48.84}\std{$\pm${1.68}} & \textbf{0.37}\std{$\pm${0.11}} &   \textbf{99.32}\std{$\pm${0.18}}  \\
\bottomrule[1.5pt]
\end{tabular*}
\end{table}

\noindent\textbf{Ablation studies.}\quad
 We start with a \generator using MSE loss and gradually insert some loss functions and our \contrablock. Table~\ref{loss_functions} presents the results of different loss functions. It shows that both the global non-contrastive module and local contrastive module are helpful in obtaining better metrics due to the capacity of exploiting inherent anatomical information and maintaining anatomical consistency.
 Then, we add our \contrablock to two supervised models: RED-CNN~\cite{redcnn} and U-Net~\cite{ronneberger2015u} but it is less effective, which demonstrates the importance of our \generator that captures both local and global contexts during denoising in Table~\ref{model_macnet}.  
 In addition, we evaluate the effectiveness of the training strategies including alternate learning, neighboring positive matching and hard negative sampling in Table~\ref{training_strategies}.

\section{Conclusion}
In this paper, we explore the anatomical semantics in LDCT denoising and take advantage of it to improve the denoising performance. To this end, we propose an \textbf{A}natomy-aware \textbf{S}upervised \textbf{CON}trastive
learning framework (\modelname), consisting of an
efficient self-attention-based U-Net (\generator) and a multi-scale anatomical contrastive network (\contrablock), which can capture both local and global contexts during denoising and exploit inherent anatomical information. 
Extensive experimental results on Mayo-2016 and Mayo-2020 datasets demonstrate the superior performance of our method, and the effectiveness of our designs. We also validated that our method introduces interpretability to LDCT denoising.

\newpage

\section*{Supplementary Materials}
\renewcommand*{\thefigure}{S\arabic{figure}}
\setcounter{figure}{0}
\renewcommand*{\thetable}{S\arabic{table}}
\setcounter{table}{0}

\begin{figure}[htbp]
\centering
\includegraphics[width=0.5\linewidth]{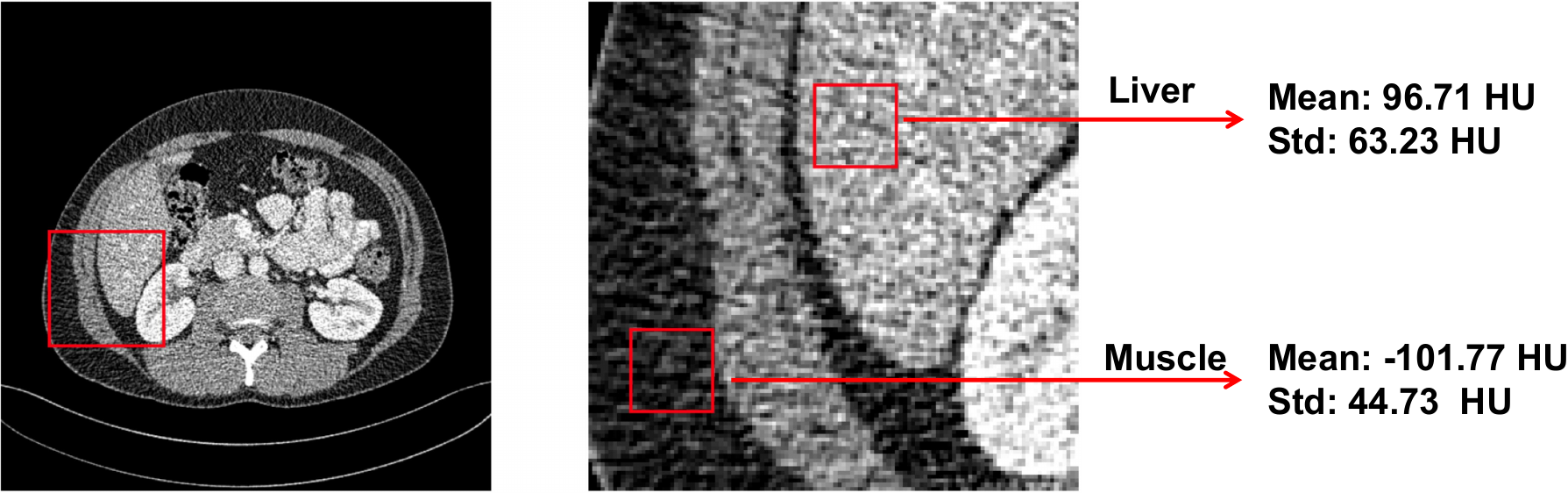}
\caption{The mean and standard deviation of different tissues in an example of an LDCT image. The standard deviation (std) of an ROI in the liver is 63.23 HU, while the std of an ROI in the muscle is 44.73 HU. }
\label{mean_std}
\end{figure}

\begin{figure*}[htbp]
\centering
\includegraphics[width=1\linewidth]{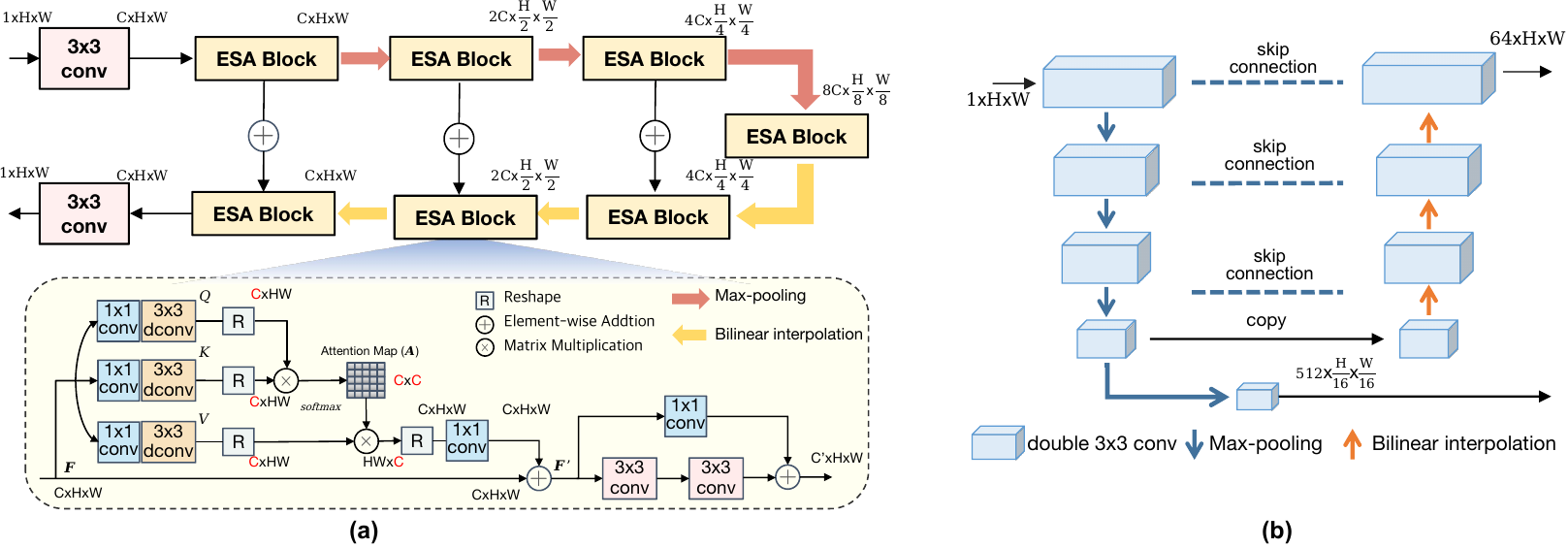}
\caption{Overview of (a) \generator and (b) disentangled U-shape architecture.}
\label{network_esau_dunet}
\end{figure*}

\begin{figure}[htbp]
\centering
\includegraphics[width=1\linewidth]{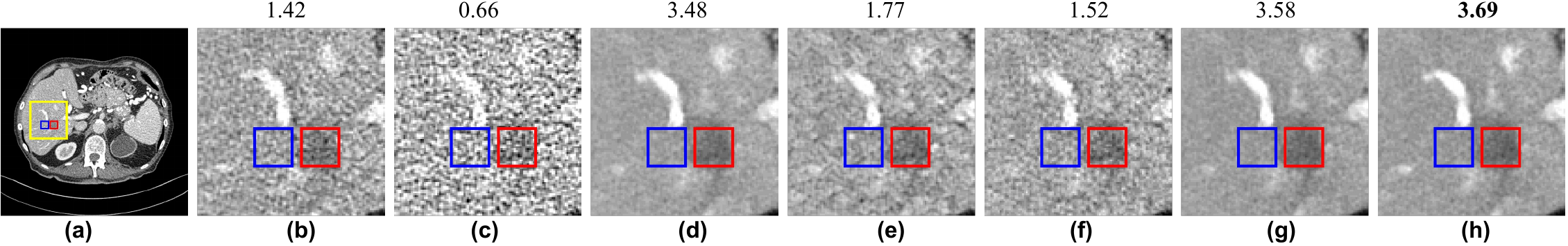}
\caption{Contrast-to-noise ratio (CNR) of the low-contrast lesion from Mayo-2016 dataset. We select the yellow rectangle of every image to visualize (see in (a)). ROIs of the red rectangle and the blue rectangle are the lesion and the background. (b) NDCT; (c) LDCT; (d) RED-CNN; (e) WGAN-VGG; (f) CNCL; (g) \generator (ours); and (h) \modelname (ours). Note that our proposed method achieves the best performance of CNR.}
\label{CNR}
\end{figure}

\begin{figure}[htbp]
\centering
\includegraphics[width=0.9\linewidth]{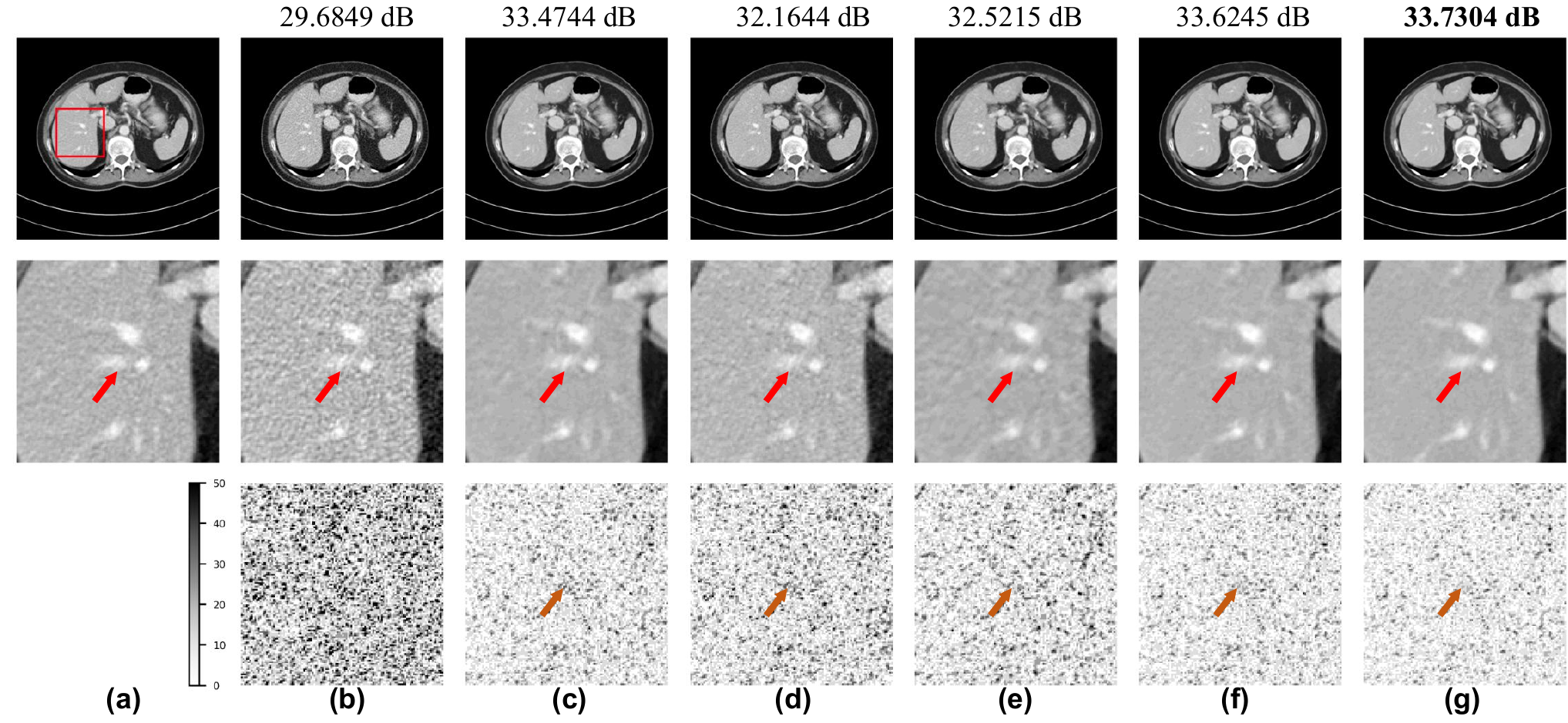}
\caption{Transverse CT images and corresponding difference images from the Mayo-2020 dataset: (a) NDCT; (b) LDCT; (c) RED-CNN; (d) WGAN-VGG; (e) CNCL; (f) \generator (ours); and (g) \modelname (ours). Zoomed ROI of the rectangle is shown below the full-size one. The display window is [-160, 240] HU.}
\label{spatial_2020}
\end{figure}

\begin{table}[b]
\centering
\renewcommand\arraystretch{0.9}
\caption{Ablation results on different supervised models with \contrablock. Note that the performance of  RED-CNN and Unet decreases when adding a \contrablock. It verifies the importance of our \generator that captures both local and global contexts during denoising.}
\resizebox{\textwidth}{!}{
\begin{tabular*}{1\linewidth}{@{\extracolsep{\fill}}ccccclccc}
\toprule
  && \multicolumn{3}{c}{\textbf{Mayo-2016 Dataset}}         && \multicolumn{3}{c}{\textbf{Mayo-2020 Dataset}} \\
\cline{3-5}  \cline{7-9}
Method && PSNR$\uparrow$ &  RMSE$\downarrow$      & SSIM$\uparrow$ && PSNR$\uparrow$   &   RMSE$\downarrow$   & SSIM$\uparrow$ \\
 \midrule
RED-CNN   && 44.23 & 0.62 & 97.34   && 48.05& 0.41 & 99.28    \\
&& {$\pm${1.26}} &  {$\pm${0.09}} & {$\pm${0.86}} && {$\pm${2.14}}   &{$\pm${0.11}}   &{$\pm${0.18}} \\

RED-CNN w/ \contrablock  && 43.55  & 0.67 &  97.11 &&  47.85 & 0.42 & 99.27 \\
&& {$\pm${1.16}} & {$\pm${0.11}} & {$\pm${0.85}} && {$\pm${2.08}} &  {$\pm${0.10}} & {$\pm${0.21}}\\

 \midrule
Unet   &&  44.13 & 0.64 &  97.38   &&     47.67  & 0.43 & 99.19 \\
&& {$\pm${1.19}} & {$\pm${0.12}} & {$\pm${1.09}} && {$\pm${1.64}} & {$\pm${0.09}}  &  {$\pm${0.23}}

\\
Unet w/ \contrablock && 44.04 & 0.65 & 97.35 &&  47.66  &  0.43 & 99.20 \\
&& {$\pm${1.15}} &  {$\pm${0.11}} &  {$\pm${0.98}}  && {$\pm${1.82}} & {$\pm${0.10}}  &{$\pm${0.21}}\\

 \midrule
\generator~ &&   44.38& 0.61 &   97.47       && 48.31   & 0.40  & 99.30\\
&& {$\pm${1.26}}  & {$\pm${0.09}}  & {$\pm${0.87}}  && {$\pm${1.87}}  &  {$\pm${0.12}} &{$\pm${0.18}}       \\

\modelname \textbf{(proposed)} && \textbf{44.48} & \textbf{0.60} & \textbf{97.49}   && \textbf{48.84} & \textbf{0.37} &   \textbf{99.32} \\

&& {$\pm${1.32}}  & {$\pm${0.10}}  &  {$\pm${0.86}}   && {$\pm${1.68}} & {$\pm${0.11}} &{$\pm${0.18}} \\
\bottomrule
\end{tabular*}}
\label{model_macnet}
\end{table}

\begin{table}[b]
\centering
\renewcommand\arraystretch{0.95}
\caption{Ablation results of Mayo-2020 on the different training strategies in terms of PSNR, RMSE, and SSIM.}
\begin{tabular*}{1\linewidth}{@{\extracolsep{\fill}}lcccc}
\toprule
Method      & PSNR $\uparrow$ & RMSE$\downarrow$ & SSIM$\uparrow$\\
 \midrule
 w/o alternate training    &  48.57{$\pm${1.82}}  & 0.38{$\pm${0.10}} & 99.28{$\pm${0.19}} \\
 w/o neighboring positive matching strategy &   48.51{$\pm${1.61}}   &   0.39{$\pm${0.10}}  &  99.31{$\pm${0.17}}   \\
w/o hard negative sampling  &  48.41{$\pm${1.73}}  &  0.40{$\pm${0.09}}  & 99.29{$\pm${0.19}} \\
\modelname \textbf{(proposed)}  &  \textbf{48.84{$\pm${1.68}}}   &  \textbf{0.37{$\pm${0.11}}}  &  \textbf{99.32{$\pm${0.18}}}  \\ 

\bottomrule
\end{tabular*}
\label{training_strategies}
\end{table}

\begin{algorithm}[htbp]
\small
\caption{Optimization of alternate learning}
\label{alternate_learning}
\begin{algorithmic}[1]
\State \textbf{Input:} LDCT $\mat{X}$, NDCT $\mat{Y}$
\State \textbf{Networks in \modelname:} \generator and \contrablock
\State \textbf{Optimizers:} Optimizer\_E (\generator) and Optimizer\_M (\contrablock)\\
\State $\mat{Y}^{\prime}$= \generator.forward($\mat{X}$)\\
\\
\# \textbf{Optimization of \contrablock}
\State Set\_requires\_grad (\contrablock, True)
\State $\operatorname{\contrablock}$ .forward($\mat{Y}^{\prime}$.detach(), $\mat{Y}$) 
\State Optimizer\_M.zero\_grad() 
\State Compute $\mathcal{L}_\mathrm{contra}=\mathcal{L}_\mathrm{local}+\mathcal{L}_\mathrm{global}$
\State$\mathcal{L}_\mathrm{contra}$.backward () 
\State Optimizer\_M.step ()\\ 
\\
\# \textbf{Optimization of \generator}
\State Set\_requires\_grad (\contrablock, False)
\State Optimizer\_E.zero\_grad () 
\State Compute $\mathcal{L}=\mathcal{L}_\mathrm{pixel}+\mathcal{L}_\mathrm{contra}$ 
\State $\mathcal{L}$.backward () 
\State Optimizer\_E.step () 
\end{algorithmic}
\end{algorithm}

\end{document}